\documentclass[conference]{IEEEtran}
\IEEEoverridecommandlockouts
\usepackage{booktabs}
\usepackage{cite}
\usepackage{amsmath,amssymb,amsfonts}
\usepackage{algorithmic}
\usepackage{graphicx}
\usepackage{textcomp}
\usepackage{xcolor}
\def\BibTeX{{\rm B\kern-.05em{\sc i\kern-.025em b}\kern-.08em
    T\kern-.1667em\lower.7ex\hbox{E}\kern-.125emX}}
\begin{document}

\title{A Novel Computing Paradigm for MobileNetV3 using Memristor}

\author{
    \IEEEauthorblockN{Jiale Li\IEEEauthorrefmark{1}\IEEEauthorrefmark{2}, Zhihang Liu\IEEEauthorrefmark{2}, Sean Longyu Ma\IEEEauthorrefmark{2}, Chiu-Wing Sham\IEEEauthorrefmark{2},Chong Fu\IEEEauthorrefmark{3}}
    \IEEEauthorblockA{\IEEEauthorrefmark{2} School of Computer Science, The University of Auckland, Auckland, New Zealand}
    \IEEEauthorblockA{\IEEEauthorrefmark{3} School of Computer Science and Engineering, Northeastern University, Shenyang, China}
    \IEEEauthorblockA{\{jli990, zliu604\}@aucklanduni.ac.nz, \{sean.ma, b.sham\}@auckland.ac.nz, fuchong@mail.neu.edu.cn}

    \thanks{\IEEEauthorrefmark{1}Corresponding Author}
}

\maketitle

\begin{abstract}
The advancement in the field of machine learning is inextricably linked with the concurrent progress in domain-specific hardware accelerators such as GPUs and TPUs. However, the rapidly growing computational demands necessitated by larger models and increased data have become a primary bottleneck in further advancing machine learning, especially in mobile and edge devices. Currently, the neuromorphic computing paradigm based on memristors presents a promising solution. In this study, we introduce a memristor-based MobileNetV3 neural network computing paradigm and provide an end-to-end framework for validation. The results demonstrate that this computing paradigm achieves over 90\% accuracy on the CIFAR-10 dataset while saving inference time and reducing energy consumption. With the successful development and verification of MobileNetV3, the potential for realizing more memristor-based neural networks using this computing paradigm and open-source framework has significantly increased. This progress sets a groundbreaking pathway for future deployment initiatives. 
\end{abstract}


\section{Introduction}

The progress in machine learning is inextricably linked to the availability of computational resources. As machine learning, particularly deep learning models, strive to enhance their task performance, the increasing demand for floating-point operations places a significant burden on computational resources~\cite{gonzalez2024spinnaker2}. Fortunately, Very Large Scale Integration (VLSI) technology has seen significant advancements~\cite{Lu2010ADA,Lu2012ANC}. Specifically, the application of graphics processing units (GPUs) has rendered the training of deep neural networks (DNNs) on large-scale datasets feasible, enabling researchers to explore more complex algorithms to augment their models~\cite{krizhevsky2012imagenet}. However, deep learning models are currently expanding towards the computational availability limits at an unprecedented rate~\cite{wozniak2020deep}, particularly in edge devices where there are stringent requirements for energy consumption and inference latency~\cite{Lo2020ANI,Yue2024MTSTAM,fu2025efficient,li2024edge}. Recently, some lightweight networks such as MobileNet~\cite{howard2019searching}, ShuffleNet~\cite{ma2018shufflenet}, and SqueezeNet~\cite{iandola2016squeezenet} have been proposed to compress the model size while maintaining high accuracy. However, even for these lightweight networks, the parameters often reach millions. Utilizing high-performance GPUs for inference at the edge is impractical due to their cost-prohibitive nature and energy-intensive characteristics~\cite{desislavov2023trends}. Moreover, other traditional hardware accelerators based on the Von Neumann architecture~\cite{Lo2020EnergyEF, Lo2018FixedPointIO, Lo2023NovelCA,yan2024survey} are unable to fundamentally resolve the issues of high energy consumption and latency caused by the extensive data movement due to the separation of memory and computing unit in the inference process of DNNs~\cite{mutlu2022modern}. As an emerging solution, neuromorphic computing based on the memristor crossbar can significantly increase speed and decrease power consumption, offering a new computing paradigm for machine
learning~\cite{zhang2023edge}.

MobileNetV3, an efficient DNN optimized for deployment on edge devices, is designed to reduce computational load without compromising recognition accuracy. Exploring how to leverage neuromorphic computing to achieve hardware-software synergy, thereby maximizing the advantages of low power consumption and rapid inference speed of MobileNetV3, is a compelling research topic. Due to the complex architecture and activation functions of MobileNetV3, research papers are scarce in this area. In this study, we introduce a new computing paradigm and hardware design for MobileNetV3 based on memristor. MobileNetV3 extensively uses vector-matrix multiplications (VMMs), which can be effectively performed in the analog domain with memristor crossbars. In this computing paradigm, multiplication is executed using Ohm's Law, and summation is achieved through Kirchhoff's Current Law~\cite{li2018analogue}. The voltage output from sensors is directly applied to the rows of a memristor crossbar. The trained weights are stored as the conductance values of the memristors. The resulting currents in each column of the array form the output vector, which is then converted to voltage signals via trans-impedance amplifiers (TIAs) for the next layer's input. Nonlinear activation functions are implemented with complementary metal-oxide-semiconductor(CMOS)-based circuits. The memristor-based MobileNetV3 network is made up of several key neural modules: memristor-based convolution module, memristor-based batch normalization module, activation function module, memristor-based global average pooling module, and memristor-based fully connected module. It also includes addition modules for residual connections~\cite{he2016deep} and multiplication modules in the attention modules~\cite{howard2019searching}. These components work together to handle all computations of MobileNetV3 using analog signals. In this approach, there is no need to move the data between the computing unit and memory, which reduces energy consumption. Additionally, the memristor-based VMM serves as an accelerator for DNN inference due to its inherent parallelism~\cite{wang2023parallel}.

To validate this computing paradigm and the hardware implementation of MobileNetV3, we also provide a framework for high-level synthesis, which converts programs written in high-level programming languages into hardware description languages~\cite{coussy2009introduction}, rapidly constructing SPICE-based netlist files. Researchers need only supply input data and weight files, and the framework automatically generates a reliable memristor-based circuit implementation based on the network structure. We used this framework to conduct image classification tests on the CIFAR-10 dataset. Results indicate that our approach surpasses other state-of-the-art memristor-based implementations in accuracy, and outperforms traditional central processing unit (CPU) and GPU-based methods in time and energy consumption. In summary, this paper makes the following main contributions:

\begin{itemize}

\item Based on memristor, a new computing paradigm and hardware design of the MobileNetV3 is presented for the first time. Experiments show that the proposed memristor-based computing paradigm achieves an accuracy of $>90\%$ on the CIFAR-10 dataset with advantages in computing resources, calculation time, and power consumption. Remarkably, this solution achieves an extraordinary acceleration, boasting a speed-up of 138 times compared to traditional GPU-based implementation and 2827 times relative to conventional CPU-based solutions. Regarding energy efficiency, it realizes savings of 4.5 times compared to GPU and 61.7 times compared to CPU.

\item Four energy-saving modular circuits with functions of layer normalization, hard sigmoid, hard swish, and convolution are firstly designed as vital components for various DNNs. Our design philosophy prioritizes simplicity, especially in the convolutional layers and fully connected layers, where we have reduced the number of operational amplifiers (op-amps) by 50\% compared to conventional implementation methods~\cite{li2022cmos,zhang2019memristive}. This strategy is intended to maximize the inherent efficiency and energy-saving benefits of our proposed computing paradigm.

\item Furthermore, we have developed an open-source, automatic mapping framework to facilitate the rapid construction and simulation of memristor-based neural networks. The code is available at: https://github.com/JialeLiLab/MMobileNetV3. This framework enables users to generate reliable netlist files within minutes, a process that would traditionally take days to complete manually~\cite{nikishov2023automated}. The framework blends flexibility with efficiency, offering researchers a powerful tool to explore new memristor-based computing paradigms in machine learning. This development not only underscores the practicality of our computing paradigm but also opens new avenues for investigation in the field.

\end{itemize}

\section{Background and Related Work}

With the surge in computational demands brought on by the development of large-scale deep learning models in machine learning, new computing paradigms founded on memristors are garnering heightened research interest~\cite{yao2020fully}. A memristor is notable for its dynamic resistance adjustment and memory of voltage or current sequences~\cite{joglekar2009elusive}. Its capabilities for synaptic emulation, efficient weight adjustment, low energy consumption, and parallel processing position it as a promising solution in new computing paradigms, particularly for neural networks and brain-inspired computing~\cite{xia2019memristive}. 

In 1971, Professor Leon Chua, guided by the principles of electronic theory, predicted the existence of the memristor~\cite{chua1971memristor}. In 2008, HP Labs provided the first experimental confirmation of Chua's hypothesis by creating a memristor device based on titanium dioxide~\cite{strukov2008missing}. Recently, neural network modules based on memristors have been developed~\cite{yakopcic2016memristor, yakopcic2017extremely, wen2020ckfo}, demonstrating the viability of a machine learning computing paradigm using memristors. However, due to the complexity of the circuits and the time-consuming nature of circuit-level simulation~\cite{xia2017mnsim}, research on the design of memristor-based classic DNNs is currently sparse and primarily focused on testing with small datasets. Specifically, Wen et al. designed a sparse and compact convolutional neural network based on memristors, obtaining a 67.21\% accuracy rate on the CIFAR-10 dataset~\cite{wen2019memristor}. Ran et al. proposed a memristor-based ShuffleNetV2 and achieved an accuracy of 55.59\% on the FER2013 dataset~\cite{ran2020memristor}. Yang et al. introduced a memristor-based Transformer Network and tested it on 3 $\times$ 3 character image recognition and 8 $\times$ 8 MNIST handwritten digit datasets~\cite{yang2022full}. In contrast, our computing paradigm is based on the widely-used DNN MobileNetV3 and achieved over 90\% accuracy on the CIFAR-10 dataset.

\section{Methodology}

\subsection{Network Architecture}
The MobileNetV3 network architecture can be divided into four main memristor-based neural network units, in which the composition and connection relationship are shown in Figure 1. The computing paradigm has the same neural network layers as~\cite{chen2021automatic}. The input layer is a preprocessing unit for input images, including a convolutional sub-layer, a normalization sub-layer, and a hard swish activation sub-layer. The body layer contains several bottleneck blocks, which include lightweight attention modules based on squeeze and excitation. The last convolutional layer is responsible for extracting high-level features from the input image and performing dimension transformation. It comprises a convolutional sub-layer, a normalization sub-layer, and a hard swish activation sub-layer. The final layer is the classification layer, featuring a global average pooling sub-layer, two fully connected sub-layers, and a hard swish activation sub-layer. The fully connected layer is instrumental in achieving the classification of the input image.
\begin{figure}[htbp]
\centerline{\includegraphics[width=0.5\textwidth]{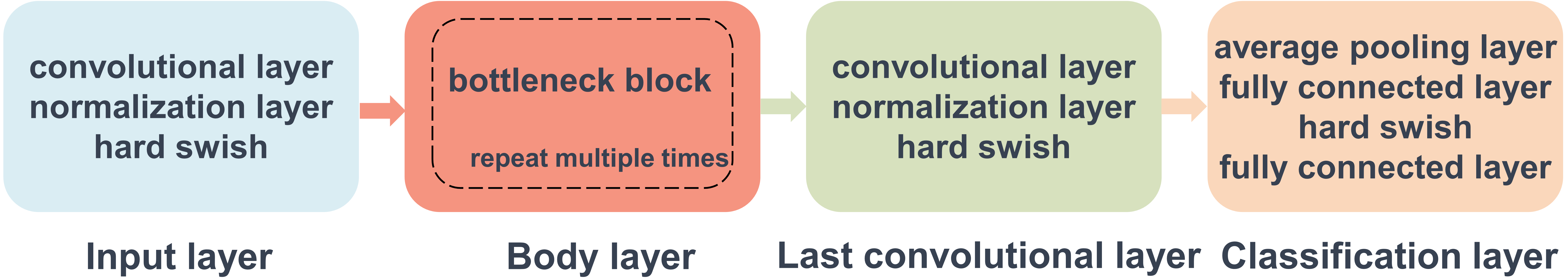}}
\caption{Flowchart of the network architecture based on MobileNetV3.}
\label{fig1}
\end{figure}

\subsection{Memristor-Based Convolution Module}
In this computing paradigm, there are three types of convolution operations: regular convolution, depthwise convolution, and pointwise convolution. The differences among these convolution operations are as follows: In regular convolution, memristors are placed at specific locations on a crossbar to perform the sliding window operation, and their output currents are interconnected for a summation function. In contrast, depthwise convolution lacks this summation operation compared to regular convolution~\cite{chollet2017xception}. Pointwise convolution is similar to one-channel regular convolution~\cite{hua2018pointwise}. Based on the analysis above, we will focus on regular convolution as a case study to present the implementation details of the memristor crossbar.
\begin{figure}[htbp]
\centerline{\includegraphics[width=0.5\textwidth]{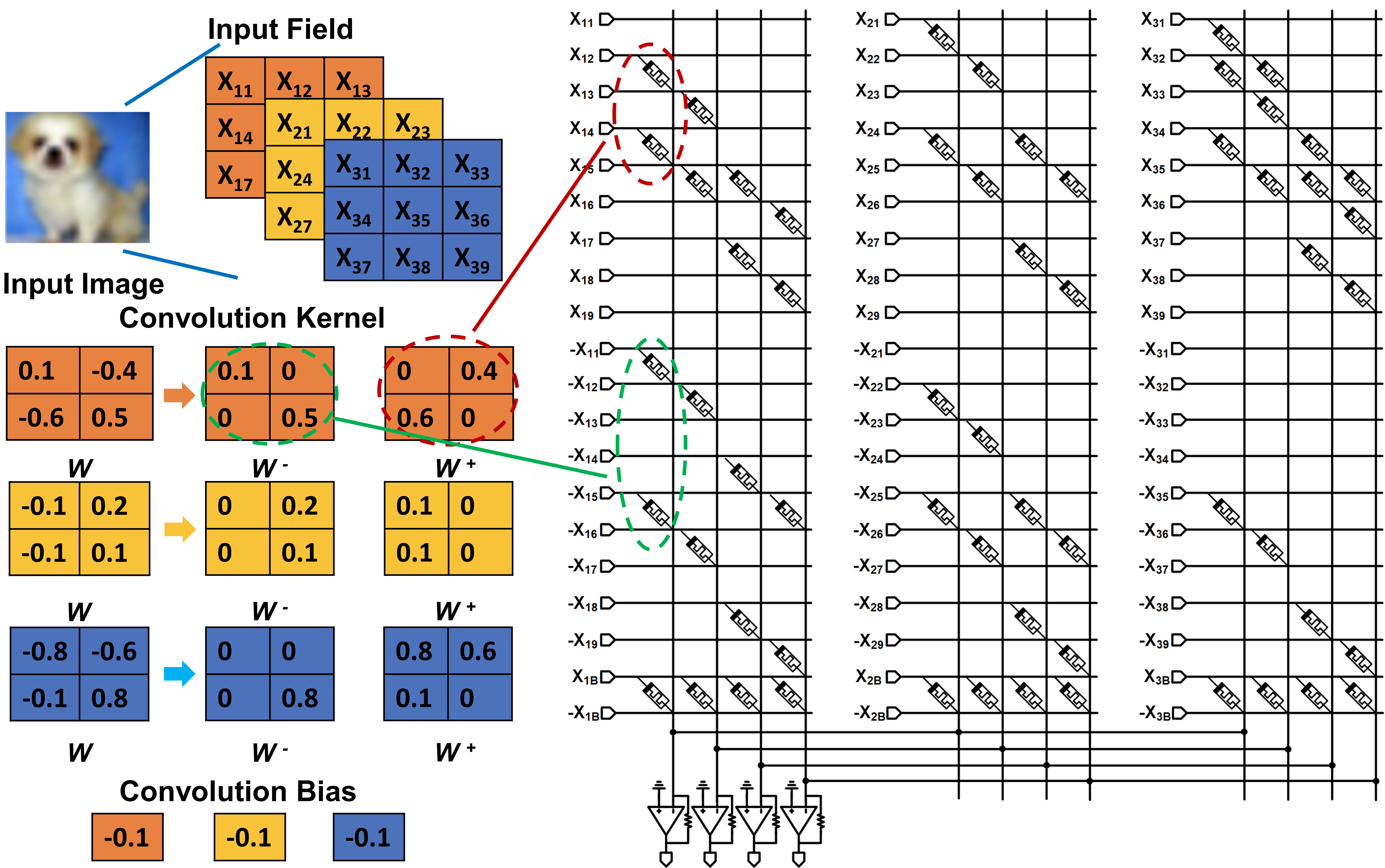}}
\caption{Circuit schematic of memristor crossbars for a convolution operation.}
\label{fig2}
\end{figure}

To perform the convolution operation between the memristor-based convolutional kernel and the input matrix, it is necessary to split the original weight matrix into positive and negative matrices firstly since the resistance of a single memristor is positive. As shown in Figure 2, contrary to the conventional approach in most research papers~\cite{yakopcic2016memristor, li2021native, yakopcic2017extremely}, we designate the matrix with positive weights as the negative weight matrix, mapping the negative weight matrix to that part of the memristor crossbar consisting of the inverting inputs, while the matrix with negative weights is labelled as the positive weight matrix, corresponding to the original inputs of the memristor crossbar. The output current has the opposite polarity to the actual result. After passing through the TIA, it is converted into a voltage with the same polarity as the actual result, due to the TIA's inverting nature. Compared to aforementioned solutions, the most significant advantage of this method is that it reduces the number of op-amps at each output port. Given that the power consumption of op-amps is at the mW level while that of memristors is at the $\mu$W level~\cite{wen2019memristor}, this approach is indeed meaningful. A more detailed analysis of power consumption will be discussed in the experimental section.

Secondly, the number of rows ($O_r$) and the number of columns ($O_c$) in the output matrix are determined by the formulas:
\begin{equation}
O_{r|c}=(\frac{W_{r|c}-F_{r|c}+2P}{S}+1)
\end{equation}
where $O_{r|c}$, $W_{r|c}$, and $F_{r|c}$ represent the number of rows or columns in the output matrix, input matrix, and convolutional kernel matrix, respectively. $P$ and $S$ stand for padding and stride, respectively.

Thirdly, the padded input matrix is utilized as the new input matrix and is unfolded row-wise to serve as the positive input. Subsequently, the negation of this new input matrix is employed as the negative input. The starting position of each output memristor in the positive and negative input regions is determined by the following equations:
\begin{equation}
P_{Pi}=(\lfloor\frac{i}{O_c}\rfloor*{W_c}+{i}\bmod{O_c})*S, i \in \mathbb{N}
\end{equation}
\begin{equation}
P_{Ni}=(\lfloor\frac{i}{O_c}\rfloor*{W_c}+{i}\bmod{O_c})*S+{W_r}*{W_c}, i \in \mathbb{N}
\end{equation}
where $i$ is the output index, and $P_{Pi}$, $P_{Ni}$ represent the starting position of $O'_i$ in the positive and negative input regions respectively. In other words, this formula indicates that ($O'_i$, $P_{Pi}$) are the coordinates for placing the first memristor in each column. Similarly, ($O'_i$, $P_{Ni}$) are the coordinates for placing the first memristor in the negative input regions of each column.

Fourthly, the convolutional kernel is unfolded into a single-row matrix. Starting from ($O'_i$, $P_{Pi}$), sequentially assign $F_c$ memristors with weights to the memristor crossbars, and repeat the process with an interval of ($W_c$-$F_c$+$2P$). It is worth noting that memristors with a weight of zero do not appear in the memristor crossbar to reduce the number of memristors, as their contribution to the output is zero. Then, starting from ($O'_i$, $P_{Ni}$), assign memristors in the negative input area according to the above rules. The two bias voltages, as the last inputs, combined with the memristors, form the bias for the convolution operation. 

Finally, the convolution operation formula in memristor crossbars is expressed by the following equation:
\begin{equation}
V_{j}=-\sum_{i=0}^{2N+1}\frac{V_{i}}{R_{i,j}}*{R_f}, i \in \mathbb{N}, j \in \mathbb{N}
\end{equation}
where ${V_{i}}$ is the $(i)^{th}$ input voltage in the memristor crossbar. $R_{i,j}$ represents the resistance of the memristor at the $(i,j)$ position, if it exists. $R_{f}$ stands for the feedback resistance in the TIA. The negative sign in the formula is determined by the properties of the TIA. ${V_{j}}$ is the $(j)^{th}$ output voltage, which corresponds to the $(j)^{th}$ element of the one-dimensional matrix obtained by flattening the convolution result matrix.

Figure 2 illustrates an example of regular convolution, where an input image is divided into three channels, each corresponding to a memristor crossbar. For regular convolution, the currents from the same output ports are aggregated and then passed through TIA. In contrast, for depthwise convolution and pointwise convolution, each output port is independently connected to TIA. The memristor array corresponding to the first channel is used as an example to illustrate the layout of memristors. The memristor crossbar consists of 20 inputs (a 3 $\times$ 3 positive matrix, a 3 $\times$ 3 negative matrix, and 2 biases) and 4 outputs (a 2 $\times$ 2 matrix) according to Formula 1. In addition, since the stride is set to one and the padding is set to zero, in this example, $O_c$, $W_c$, and $F_c$ are two, three, and two, respectively. The initial position of the memristors for each convolutional output can be determined based on Formula 2 (1 for $O'_0$, 2 for $O'_1$, 4 for $O'_2$, and 5 for $O'_3$).Then, place memristors with weights of the convolutional kernel’s first-row elements (0, 0.4) at the starting position in the memristor crossbar sequentially. After leaving a gap of one (according to $W_c$-$F_c$+$2P$), continue placing two memristors with weights of the second-row elements (0.6,0). Since there are two memristors with weights of zero, only two memristors with respective weights of 0.4 and 0.6 are placed in the positive input area. The initial position of the memristors for negative input region of each convolutional output can be determined based on Formula 3 (9 for $O'_0$, 10 for $O'_1$, 12 for $O'_2$, and 13 for $O'_3$). Then, place memristors with weights of the convolutional kernel’s first-row elements (0.1, 0) at the starting position in the memristor crossbar sequentially. After leaving a gap of one (according to $W_c$-$F_c$+$2P$), continue placing two memristors with weights of the second-row elements (0,0.5). Consistent with the previous analysis, only two memristors are placed on the negative input region of the memristor crossbar. Since the convolution bias is negative in this example, the memristor will be combined with a positive bias voltage to form the bias. In the design of the regular convolutional layer based on memristors, the number of memristors and op-amps required is given by the following formulas, respectively:
\begin{equation}
N_{cm}={O_c}*{O_r}*(F_r*F_c*{C_i}+1)*{C_o}
\end{equation}
\begin{equation}
N_{co}={O_c}*{O_r}*{C_o}
\end{equation}
In these equations,  ${C_i}$ and ${C_o}$ represent the number of input and output channels, respectively, while ${N_{cm}}$ and ${N_{co}}$ denote the number of memristors and op-amps in the convolutional layer.

\subsection{Memristor-Based Batch Normalization Module}
The batch normalization module is extensively utilized in the MobileNetV3 network architecture to achieve both excellent accuracy and efficiency. The calculation method is as follows:
\begin{equation}
y=\frac{x-E[x]}{\sqrt{Var[x]+\epsilon}}*\gamma+\beta
\end{equation}
where $x$ represents the original feature value, $E[x]$ is the mean, and $Var[x]$ is the variance, with epsilon serving as a small constant to prevent division by zero (refer to~\cite{ioffe2015batch}). $\gamma$ and $\beta$ are the learnable scaling and shifting parameters, respectively. $y$ is the output of the batch normalization module. Owing to the applicability of memristor crossbars for multiplication and addition computations, the batch normalization formula is segmented into three components: subtraction, multiplication, and addition operations. The converted formulas are as follows:
\begin{equation}
y={(x-E[x])}*\left| \frac{\gamma}{\sqrt{Var[x]+\epsilon}} \right|+\beta,\gamma\geq0
\end{equation}
\begin{equation}
y={(E[x]-x)}*\left| \frac{\gamma}{\sqrt{Var[x]+\epsilon}} \right|+\beta,\gamma<0
\end{equation}
In this circuit design, memristors and TIAs are employed to execute the batch normalization operation. Specifically, the subtraction part comprises four input ports and two memristors, with the output converted to voltage via TIA for subsequent multiplication operation. Addition is achieved like biasing in convolution operations. To minimize the use of op-amps in the design of the batch normalization module, a set of specific rules is followed. $V_b$ is set to a constant value of one. As shown in Figure 3(a), when the input value $\gamma$ is non-negative, the resistance values of the first set of memristors are sequentially set to (1, 0, 0, 1), facilitating the subtraction operation $(x-E[x])$. The resistance values of the second set of memristors depend on the sign of $\beta$. For positive $\beta$, the values are set to ($\left| \frac{\gamma}{\sqrt{Var[x]+\epsilon}} \right|$, 0, $\frac{1}{\beta}$); for negative $\beta$, they are set to ($\left| \frac{\gamma}{\sqrt{Var[x]+\epsilon}} \right|$, $\frac{1}{\beta}$, 0). As shown in Figure 3(b), in cases where $\gamma$ is negative, the first set of memristors is set to (0, 1, 1, 0) to achieve the subtraction operation $(E[x]-x)$, while the second set follows the same resistance value assignment as in the non-negative case. This approach effectively reduces the need for op-amps and enhances the overall efficiency of the batch normalization module. In the design of the batch normalization module based on memristors, the number of memristors and op-amps required are given by the following formulas, respectively:
\begin{equation}
N_{bm}=4*{C}
\end{equation}
\begin{equation}
N_{bo}=2*{C}
\end{equation}
where $C$ represents the number of channels, while $N_{bm}$ and $N_{bo}$ denote the number of memristors and op-amps in the batch normalization module.
\begin{figure}[htbp]
\centerline{\includegraphics[width=0.5\textwidth]{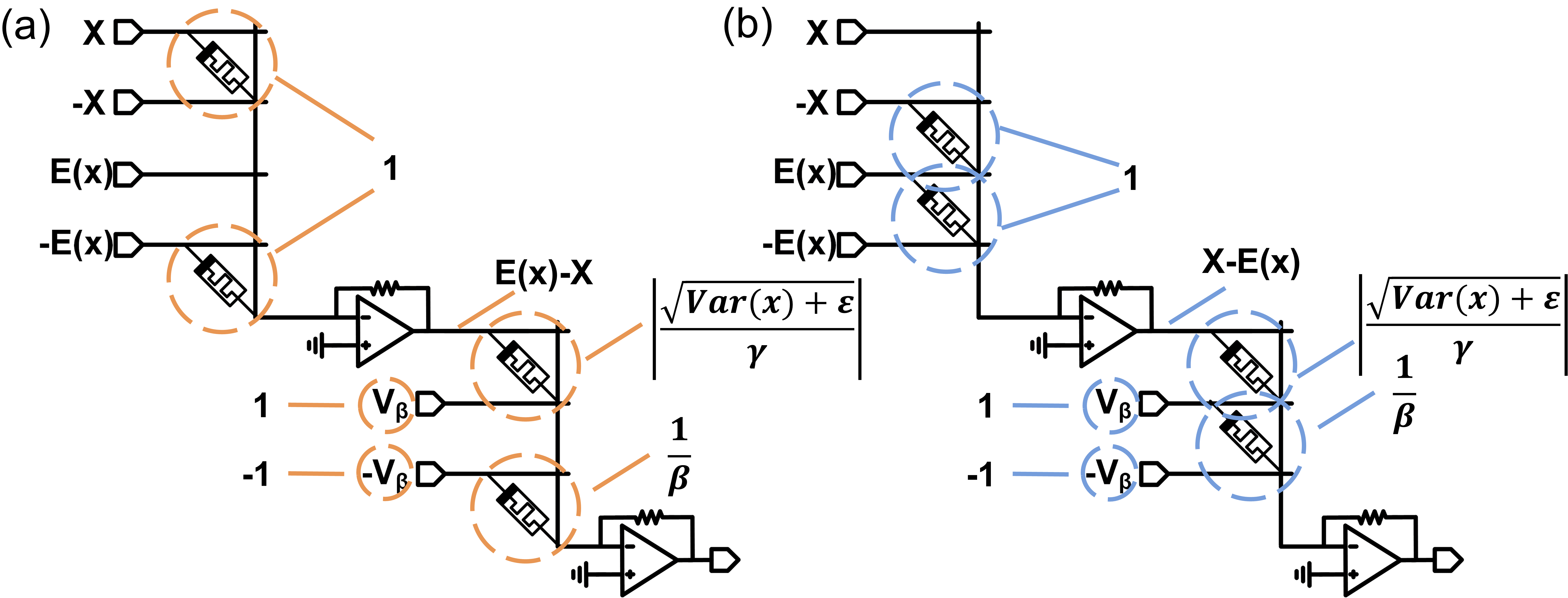}}
\caption{Circuit schematic of memristor crossbars for a batch normalization operation. (a) $\gamma\geq0$, $\beta>0$ (b) $\gamma<0$, $\beta<0$.}
\label{fig3}
\end{figure}

\subsection{Activation Function Module}
In MobileNetV3, three types of activation functions are utilized: ReLU, hard sigmoid, and hard swish. The implementation of ReLU in this paper is based on the paper published before~\cite{priyanka2019cmos}. Furthermore, we designed the hard sigmoid and hard swish activation function circuits for the first time, as shown in Figures 4(a) and 4(b). In the circuit implementation, op-amps are utilized to carry out addition and division operations. The limiter, constructed from a diode and a power source, serves the crucial role of executing the maximization function. Compared with the hard sigmoid activation function module, the hard swish module has an additional multiplication operation consisting of a multiplier. According to the simulation results shown in Figures 4(c) and 4(d), it is evident that this module achieves the functional objectives consistent with those of the software design.
\begin{figure}[htbp]
\centerline{\includegraphics[width=0.5\textwidth]{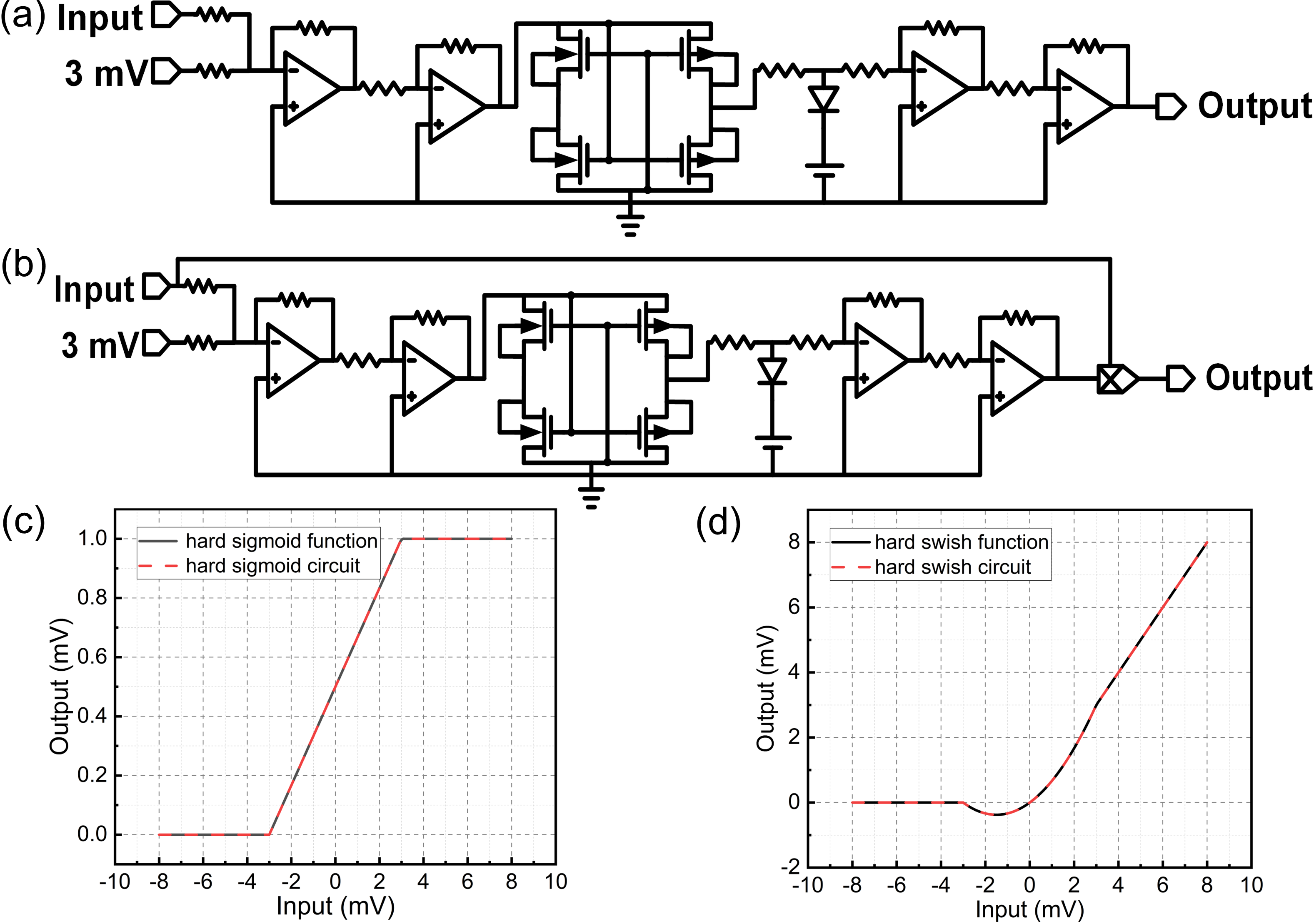}}
\caption{Circuit schematic of an activation function operation. (a) hard sigmoid, (b) hard swish, and simulation result (c) hard sigmoid, (d) hard swish.}
\label{fig4}
\end{figure}
\subsection{Memristor-Based Global Average Pooling Module}
In MobileNetV3, the global average pooling is positioned at the terminus of the network, where it transforms the deep convolutional feature maps into a one-dimensional feature vector. This resultant feature vector is subsequently utilized for classification tasks. Figure 5(a) illustrates the process of performing global average pooling computations in the memristor crossbars. The inverse of the input matrix is unfolded into a one-dimensional vector and applied as voltage to the input terminals of the memristor crossbar. The weight values are set equal to the number of input matrices. Following Ohm's Law and Kirchhoff's Circuit Laws, each input is divided by the total number of inputs and then summed, leading to a negative global average current. This negative global average current is converted into a positive global average voltage by TIAs, which serves as the output result. In the design of the global average pooling module based on memristors, the number of memristors and op-amps required are given by the following formulas, respectively:
\begin{equation}
N_{gm}={W_c}*{W_r}*{C}
\end{equation}
\begin{equation}
N_{go}={C}
\end{equation}
where $N_{gm}$ and $N_{go}$ denote the number of memristors and op-amps in the global average pooling module.
\subsection{Memristor-Based Fully Connected Module}
\begin{figure}[htbp]
\centerline{\includegraphics[width=0.5\textwidth]{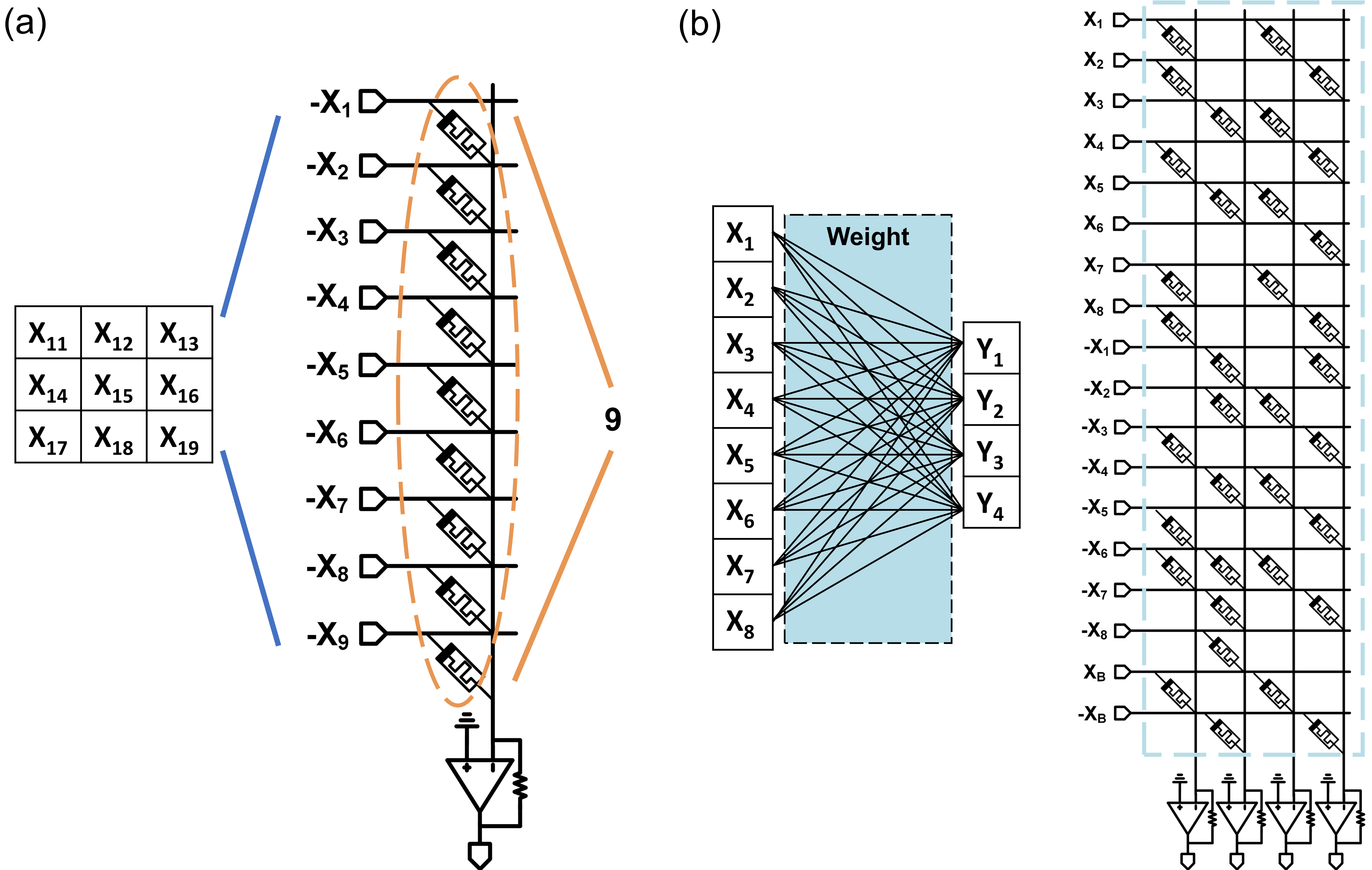}}
\caption{Circuit schematic of memristor crossbars for (a) a global average pooling operation, (b)  a fully connected operation.}
\label{fig5}
\end{figure}
The memristor-based fully connected module integrates and normalizes the highly abstracted features after multiple convolutions and outputs a probability for each category of the recognition network. The circuit of the memristor-based fully connected module is similar to that of the memristor-based convolution module, but the size is generally larger and contains more memristors. Compared to convolution modules, the arrangement rules for memristors in fully connected modules are relatively simple. It only requires converting the original weight matrix into positive and negative weight matrices, and then arranging them in a vertical sequence. In the design of the fully connected module based on memristors, the number of memristors and op-amps required are given by the following formulas, respectively:
\begin{equation}
N_{fm}={(W+1)}*{O}
\end{equation}
\begin{equation}
N_{fo}={O}
\end{equation}
where $W$ is the number of inputs. $O$ denotes the number of outputs. $N_{fm}$ and $N_{fo}$ denote the number of memristors and op-amps in the fully connected module.
\subsection{Automated Framework}
\begin{figure}[h]
\centerline{\includegraphics[width=0.5\textwidth]{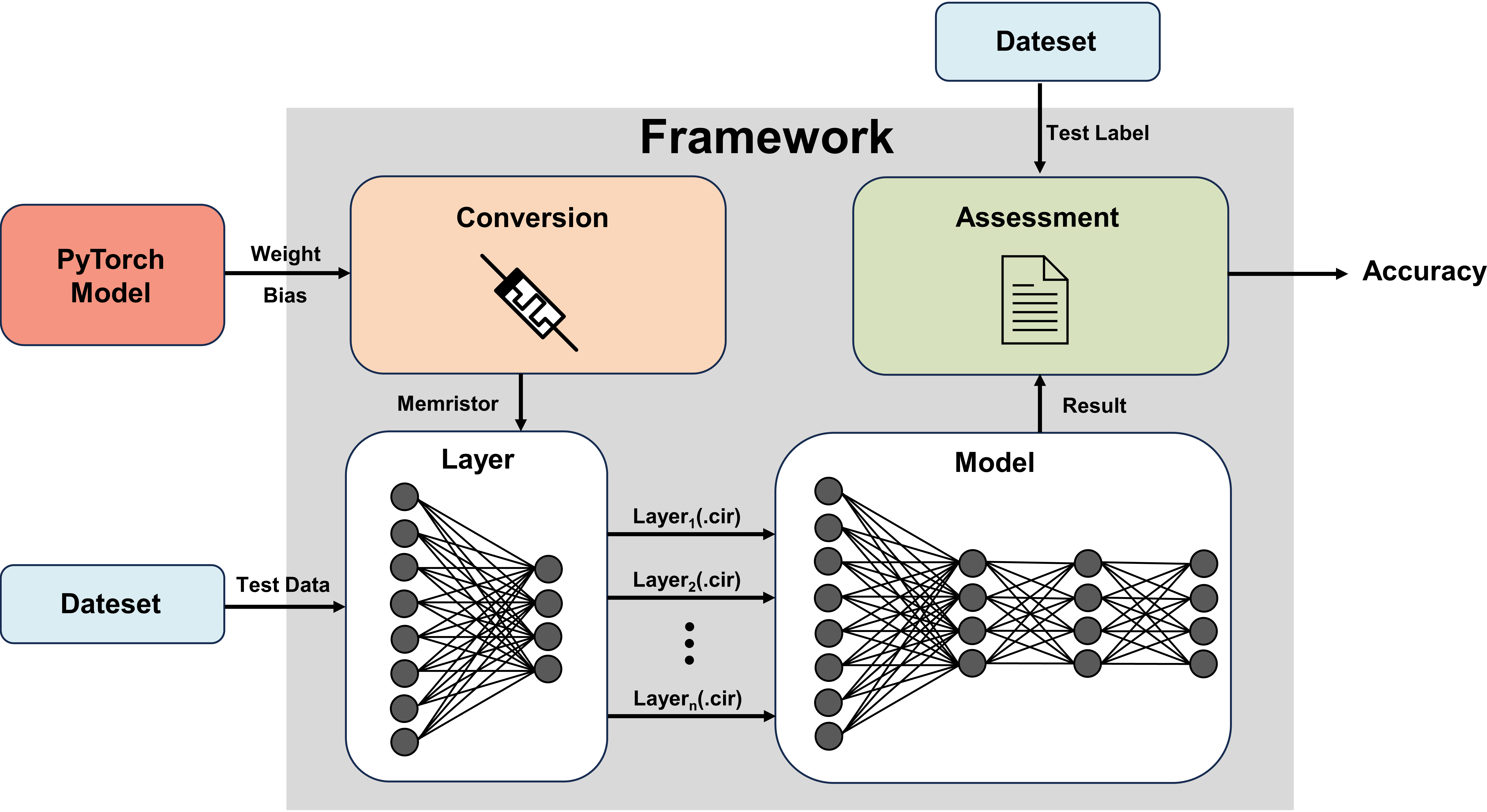}}
\caption{Block diagram of the automated framework.}
\label{fig6}
\end{figure}
Although the memristor-based computing paradigm demonstrates high energy efficiency in performing neuromorphic computations, offering a promising solution to address the limits of computational feasibility in machine learning, the development in this field is still hindered by the complexity of circuit structures and the extensive simulation analysis required as the number of neural network layers increases~\cite{li2024framework}. To solve the above problems and verify our proposed computing paradigm, we present an automated mapping framework designed for the rapid construction of the memristor-based MobileNetV3 neural network. Figure 6 shows the block diagram of the automated framework that generates the SPICE netlist of memristor circuits based on the network topology and specific hyperparameters. The framework consists of four main modules: conversion module, layer module, model module, and assessment module. The conversion module is responsible for converting the weights and biases trained with PyTorch into the format required for the resistance of the memristor. In this paper, the HP model is used to build the MobileNetV3 neural network and follows the following formula~\cite{li2021native}:
\begin{equation}
R_M=R_{on}w+R_{off}(1-w)
\end{equation}
where $R_M$, $R_{on}$, and $R_{off}$ represent the resistance of the memristor, the doped layer, and the undoped layer, respectively. $w$ is the normalized width of the doped layer. The weight from the model is used as $\frac{1}{R_M}$ to calculate the required parameter $w$ of memristor using Formula 16. The layer module generates SPICE circuit netlist files for memristor-based circuits based on the rules mentioned in the previous section for different layers. After generating the memristor-based neural network, this framework automatically performs image classification tasks on the developed memristor-based circuit by running SPICE simulations, using datasets and weight files provided by the user. The framework facilitates the creation of netlist files with second-level latency, thus opening avenues for researchers to map DNNs onto memristor crossbars. The capability to generate netlist files for large-scale memristor crossbars in a remarkably short duration is a result of abstracting the layout design rules of memristors into an algorithm. Firstly, for the memristor-based convolutional module, we implement the layout method using multiple loops, which allows for efficient organization of the memristors in the crossbar. Moving on, in the memristor-based batch normalization module, the layout of memristors is determined by conditional statements based on the sign of $\gamma$ and $\beta$. Next, the memristor-based global average pooling module uses the number of inputs as a key factor to determine the resistance values of memristors. Lastly, for the memristor-based fully connected module, we arrange the positive and negative weight matrices in a vertical sequence within the memristor crossbar.

\section{Experiments and Results}
\subsection{Accuracy}
\begin{table}[t]
\caption{Performance comparison with other neural networks based on new computing paradigms}
\label{sample-table}
\vskip 0.1in
\begin{center}
\begin{small}
\begin{tabular}{cccc}
\toprule
Publication/Year & Device & Signal & Accuracy \\
\midrule
DATE'18~\cite{sun2018xnor}    & RRAM& Digital& 86.08\% \\
TNSE'19~\cite{wen2019memristor} & memristor& Analog& 67.21\%\\
TNNLS'20~\cite{ran2020memristor}    & memristor& Analog& 84.38\%\\
ISSCC’21~\cite{xie202116}    & eDRAM& Analog& 80.1\%\\
TCASII’23~\cite{li2023adc}      & RRAM& Digital& 86.2\% \\
TCASII’23~\cite{xiao2023efficient}    & memristor& Analog& 87.5\%\\
\textbf{This work} & memristor& Analog& \textbf{90.36\%}\\
\bottomrule
\end{tabular}
\end{small}
\end{center}
\vskip -0.1in
\end{table}
In this section, we employ our developed framework to implement a scaled-down MobileNetV3 neural network for a classification application using the CIFAR-10 dataset, which comprises 32$\times$32 colour images. Following neural network computation, the CIFAR-10 dataset is categorized into ten distinct labels. The classification result determined by the memristor-based neural network is identified as the output with the highest current. The network weights are obtained from an offline server that was trained on the CIFAR-10 dataset. Table 1 compares the performance of our study with other neural networks utilizing novel computing paradigms. Our memristor-based MobileNetV3 neural network achieves a high classification accuracy of 90.36\% on the CIFAR-10 dataset, comparable to traditional implementations using the PyTorch framework and exhibiting a significant improvement over previous works in this area.
\subsection{Latency}
\begin{figure}[htbp]
\centerline{\includegraphics[width=0.5\textwidth]{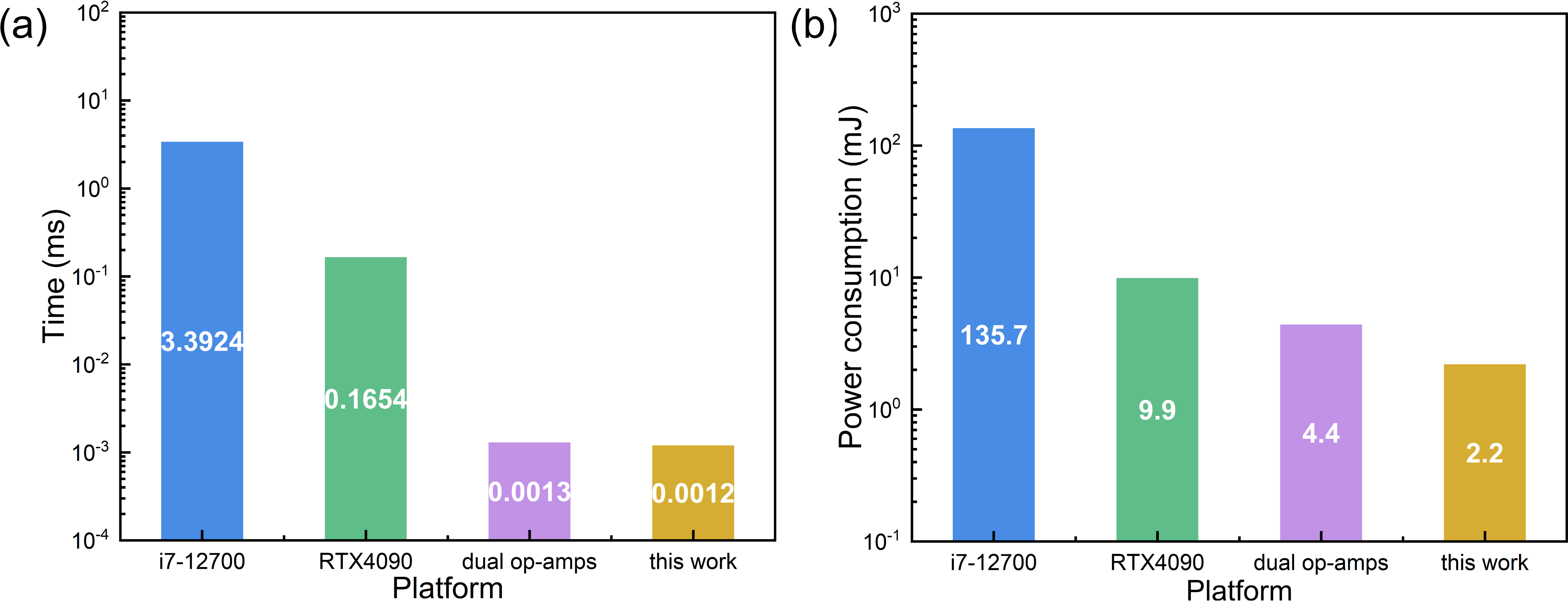}}
\caption{(a) Latency and (b) power consumption of memristor-based MobileNetV3 neural network.}
\label{fig8}
\end{figure}
We refer to methods previously reported in the papers to analyze the latency and energy consumption of memristor-based neural networks~\cite{wen2019memristor, ran2020memristor, yang2022full, zhang2023edge}, to determine the latency of our proposed computing paradigm. In neural circuits based on memristors, the response time can be as quick as 100 ps~\cite{ran2020memristor}. However, the overall speed of the memristor-based circuit is constrained by the slew rate of the op-amps employed for converting current to voltage. So the latency of this network follows the following formula:
\begin{equation}
T_i=(T_m+T_o)*N_m+T_r
\end{equation}
where $T_m$ stands for the response time of the memristor crossbar, $T_o$ is the transition time of op-amps, $N_m$ denotes the number of memristor-based layers, $T_r$ represents the latency of other layers, and $T_i$ represents the latency of the inference process. The typical slew rate of the low-power op-amps is at 10 V/$\mu$s level~\cite{hosticka1989very}. $T_r$ includes the delays associated with modules such as the activation function layer, adders, and multipliers. We analyze the latency of these layers using information on existing devices. Therefore, the latency of this circuit can be as low as 1.24 $\mu$s, in comparison to 1.30 $\mu$s for the traditional dual op-amp solution. This is significantly faster than the traditional GPU approach, which has a latency of 0.1654 ms on a server equipped with an RTX 4090. As depicted in Figure 7(a), we also conducted inference time tests for single-image processing on a CPU (i7-12700), resulting in a latency of 3.3924 ms. To sum up, this computing paradigm achieves an extraordinary acceleration, boasting a speed-up of over 138 times compared to traditional GPU-based implementation and over 2827 times relative to conventional CPU-based solutions.
\subsection{Energy Consumption}
Figure 8 illustrates the distribution of memristor weights across different layers. As observed in Figure 8, the weights of the memristor crossbars predominantly range between -0.2 and 0.2. The power consumption of the memristor-based neural network is derived from the weights of the memristors. Since we have mapped the input data to ±2.5 mV, we can estimate the maximum predicted power consumption of a single memristor by assuming that all input data is at 2.5 mV and all weights are 0.2. Therefore, the maximum power consumption of memristors can be estimated at 1.1 $\mu$W. Consequently, we use the following formula to estimate the energy consumption:
\begin{equation}
W_i=\sum U_{max}^2G_{max}*T_m+P_{o}*T_o+P{r}*T_r
\end{equation}
where $U_{max}$ denotes the maximum voltage across a single memristor,$G_{max}$ refers to the equivalent conductance of the memristors, $P_o$ represents the power consumption of op-amps, $P_r$ indicates the power used by other layers, and $W_i$ signifies the total power consumption. Based on the standard power consumption of devices, we estimate that a single complete forward inference of the neural network consumes 2.2 mJ. The comparative result of implementations using CPU, GPU, and dual op-amps is illustrated in Figure 7(b). As demonstrated in the figure, it achieves energy savings of 4.5 times compared to GPU implementation and 61.7 times compared to CPU implementation.
\begin{figure}[htbp]
\centerline{\includegraphics[width=0.5\textwidth]{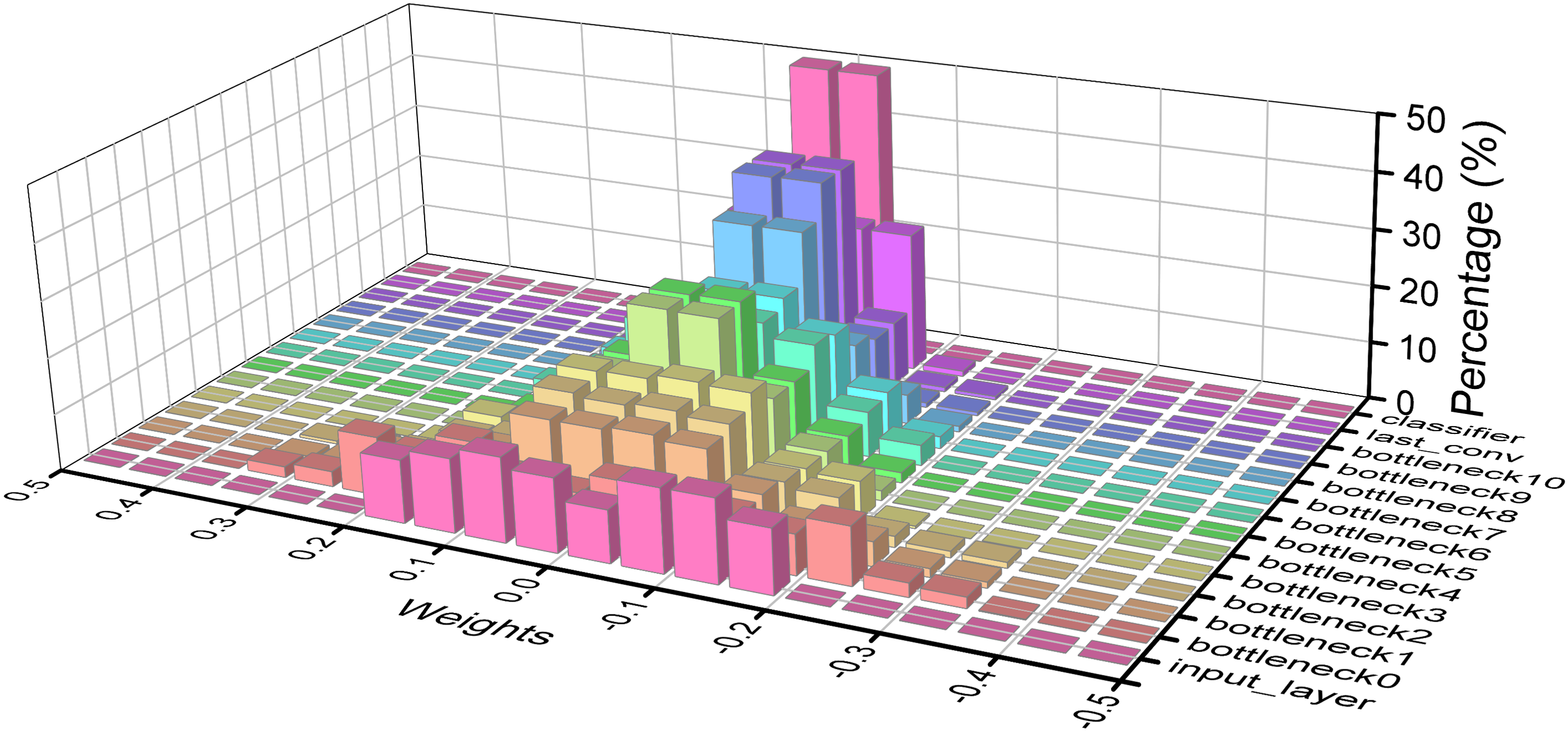}}
\caption{Distribution of weights of memristors.}
\label{fig9}
\end{figure}

\section{Conclusion}
The memristor-based MobileNetV3 proposed in this paper demonstrates exceptional performance in terms of accuracy, latency, and energy consumption, underscoring the potential of this computing paradigm as a promising solution to overcome the computational resource constraints currently faced in machine learning. Additionally, the presentation of an automated framework for constructing and validating the memristor-based computing paradigm is poised to accelerate research in this field. Our results demonstrate that the approach we propose holds advantages not only compared to traditional CPU and GPU implementations but also against other novel computing paradigms, thus paving the way for the application of memristor-based computing paradigms in the realm of machine learning. 

\bibliographystyle{IEEEtran}
\bibliography{reference}

\begin{thebibliography}{10}
\providecommand{\url}[1]{#1}
\csname url@samestyle\endcsname
\providecommand{\newblock}{\relax}
\providecommand{\bibinfo}[2]{#2}
\providecommand{\BIBentrySTDinterwordspacing}{\spaceskip=0pt\relax}
\providecommand{\BIBentryALTinterwordstretchfactor}{4}
\providecommand{\BIBentryALTinterwordspacing}{\spaceskip=\fontdimen2\font plus
\BIBentryALTinterwordstretchfactor\fontdimen3\font minus \fontdimen4\font\relax}
\providecommand{\BIBforeignlanguage}[2]{{%
\expandafter\ifx\csname l@#1\endcsname\relax
\typeout{** WARNING: IEEEtran.bst: No hyphenation pattern has been}%
\typeout{** loaded for the language `#1'. Using the pattern for}%
\typeout{** the default language instead.}%
\else
\language=\csname l@#1\endcsname
\fi
#2}}
\providecommand{\BIBdecl}{\relax}
\BIBdecl

\bibitem{gonzalez2024spinnaker2}
H.~A. Gonzalez, J.~Huang, F.~Kelber, K.~K. Nazeer, T.~Langer, C.~Liu, M.~Lohrmann, A.~Rostami, M.~Sch{\"o}ne, B.~Vogginger \emph{et~al.}, ``Spinnaker2: A large-scale neuromorphic system for event-based and asynchronous machine learning,'' \emph{arXiv preprint arXiv:2401.04491}, 2024.

\bibitem{Lu2010ADA}
J.~Lu, W.-K. Chow, C.-W. Sham, and E.~F.~Y. Young, ``A dual-mst approach for clock network synthesis,'' \emph{2010 15th Asia and South Pacific Design Automation Conference (ASP-DAC)}, pp. 467--473, 2010.

\bibitem{Lu2012ANC}
J.~Lu, W.-K. Chow, and C.-W. Sham, ``A new clock network synthesizer for modern vlsi designs,'' \emph{Integr.}, vol.~45, pp. 121--131, 2012.

\bibitem{krizhevsky2012imagenet}
A.~Krizhevsky, I.~Sutskever, and G.~E. Hinton, ``Imagenet classification with deep convolutional neural networks,'' \emph{Advances in neural information processing systems}, vol.~25, 2012.

\bibitem{wozniak2020deep}
S.~Wo{\'z}niak, A.~Pantazi, T.~Bohnstingl, and E.~Eleftheriou, ``Deep learning incorporating biologically inspired neural dynamics and in-memory computing,'' \emph{Nature Machine Intelligence}, vol.~2, no.~6, pp. 325--336, 2020.

\bibitem{Lo2020ANI}
C.~Y. Lo, C.-W. Sham, and L.~Ma, ``A novel iris verification framework using machine learning algorithm on embedded systems,'' \emph{2020 IEEE 9th Global Conference on Consumer Electronics (GCCE)}, pp. 173--175, 2020.

\bibitem{Yue2024MTSTAM}
Z.~Yue, D.~Yan, R.~Wu, L.~Ma, and C.-W. Sham, ``Mtst: A multi-task scheduling transformer accelerator for edge computing,'' \emph{2024 IEEE 13th Global Conference on Consumer Electronics (GCCE)}, pp. 1394--1395, 2024.

\bibitem{fu2025efficient}
Y.~Fu, D.~Yan, J.~Li, S.~L. Ma, C.-W. Sham, and H.-f. Chou, ``An efficient fpga-based edge ai system for railway fault detection,'' \emph{IEEE Consumer Electronics Magazine}, 2025.

\bibitem{li2024edge}
J.~Li, Y.~Fu, D.~Yan, S.~L. Ma, and C.-W. Sham, ``An edge ai system based on fpga platform for railway fault detection,'' in \emph{2024 IEEE 13th Global Conference on Consumer Electronics (GCCE)}.\hskip 1em plus 0.5em minus 0.4em\relax IEEE, 2024, pp. 1387--1389.

\bibitem{howard2019searching}
A.~Howard, M.~Sandler, G.~Chu, L.-C. Chen, B.~Chen, M.~Tan, W.~Wang, Y.~Zhu, R.~Pang, V.~Vasudevan \emph{et~al.}, ``Searching for mobilenetv3,'' in \emph{Proceedings of the IEEE/CVF international conference on computer vision}, 2019, pp. 1314--1324.

\bibitem{ma2018shufflenet}
N.~Ma, X.~Zhang, H.-T. Zheng, and J.~Sun, ``Shufflenet v2: Practical guidelines for efficient cnn architecture design,'' in \emph{Proceedings of the European conference on computer vision (ECCV)}, 2018, pp. 116--131.

\bibitem{iandola2016squeezenet}
F.~N. Iandola, S.~Han, M.~W. Moskewicz, K.~Ashraf, W.~J. Dally, and K.~Keutzer, ``Squeezenet: Alexnet-level accuracy with 50x fewer parameters and< 0.5 mb model size,'' \emph{arXiv preprint arXiv:1602.07360}, 2016.

\bibitem{desislavov2023trends}
R.~Desislavov, F.~Mart{\'\i}nez-Plumed, and J.~Hern{\'a}ndez-Orallo, ``Trends in ai inference energy consumption: Beyond the performance-vs-parameter laws of deep learning,'' \emph{Sustainable Computing: Informatics and Systems}, vol.~38, p. 100857, 2023.

\bibitem{Lo2020EnergyEF}
C.~Y. Lo and C.-W. Sham, ``Energy efficient fixed-point inference system of convolutional neural network,'' \emph{2020 IEEE 63rd International Midwest Symposium on Circuits and Systems (MWSCAS)}, pp. 403--406, 2020.

\bibitem{Lo2018FixedPointIO}
C.~Y. Lo, F.~C.-M. Lau, and C.-W. Sham, ``Fixed-point implementation of convolutional neural networks for image classification,'' \emph{2018 International Conference on Advanced Technologies for Communications (ATC)}, pp. 105--109, 2018.

\bibitem{Lo2023NovelCA}
C.~Y. Lo, C.-W. Sham, and C.~Fu, ``Novel cnn accelerator design with dual benes network architecture,'' \emph{IEEE Access}, vol.~11, pp. 59\,524--59\,529, 2023.

\bibitem{yan2024survey}
F.~Yan, A.~Koch, and O.~Sinnen, ``A survey on fpga-based accelerator for ml models,'' \emph{arXiv preprint arXiv:2412.15666}, 2024.

\bibitem{mutlu2022modern}
O.~Mutlu, S.~Ghose, J.~G{\'o}mez-Luna, and R.~Ausavarungnirun, ``A modern primer on processing in memory,'' in \emph{Emerging Computing: From Devices to Systems: Looking Beyond Moore and Von Neumann}.\hskip 1em plus 0.5em minus 0.4em\relax Springer, 2022, pp. 171--243.

\bibitem{zhang2023edge}
W.~Zhang, P.~Yao, B.~Gao, Q.~Liu, D.~Wu, Q.~Zhang, Y.~Li, Q.~Qin, J.~Li, Z.~Zhu \emph{et~al.}, ``Edge learning using a fully integrated neuro-inspired memristor chip,'' \emph{Science}, vol. 381, no. 6663, pp. 1205--1211, 2023.

\bibitem{li2018analogue}
C.~Li, M.~Hu, Y.~Li, H.~Jiang, N.~Ge, E.~Montgomery, J.~Zhang, W.~Song, N.~D{\'a}vila, C.~E. Graves \emph{et~al.}, ``Analogue signal and image processing with large memristor crossbars,'' \emph{Nature electronics}, vol.~1, no.~1, pp. 52--59, 2018.

\bibitem{he2016deep}
K.~He, X.~Zhang, S.~Ren, and J.~Sun, ``Deep residual learning for image recognition,'' in \emph{Proceedings of the IEEE conference on computer vision and pattern recognition}, 2016, pp. 770--778.

\bibitem{wang2023parallel}
C.~Wang, G.-J. Ruan, Z.-Z. Yang, X.-J. Yangdong, Y.~Li, L.~Wu, Y.~Ge, Y.~Zhao, C.~Pan, W.~Wei \emph{et~al.}, ``Parallel in-memory wireless computing,'' \emph{Nature Electronics}, pp. 1--9, 2023.

\bibitem{coussy2009introduction}
P.~Coussy, D.~D. Gajski, M.~Meredith, and A.~Takach, ``An introduction to high-level synthesis,'' \emph{IEEE Design \& Test of Computers}, vol.~26, no.~4, pp. 8--17, 2009.

\bibitem{li2022cmos}
B.~Li and G.~Shi, ``A cmos rectified linear unit operating in weak inversion for memristive neuromorphic circuits,'' \emph{Integration}, vol.~87, pp. 24--28, 2022.

\bibitem{zhang2019memristive}
Y.~Zhang, M.~Cui, L.~Shen, and Z.~Zeng, ``Memristive quantized neural networks: A novel approach to accelerate deep learning on-chip,'' \emph{IEEE transactions on cybernetics}, vol.~51, no.~4, pp. 1875--1887, 2019.

\bibitem{nikishov2023automated}
D.~Nikishov and A.~Antonov, ``Automated generation of spice models of memristor-based neural networks from python models,'' in \emph{2023 International Conference on Industrial Engineering, Applications and Manufacturing (ICIEAM)}.\hskip 1em plus 0.5em minus 0.4em\relax IEEE, 2023, pp. 895--899.

\bibitem{yao2020fully}
P.~Yao, H.~Wu, B.~Gao, J.~Tang, Q.~Zhang, W.~Zhang, J.~J. Yang, and H.~Qian, ``Fully hardware-implemented memristor convolutional neural network,'' \emph{Nature}, vol. 577, no. 7792, pp. 641--646, 2020.

\bibitem{joglekar2009elusive}
Y.~N. Joglekar and S.~J. Wolf, ``The elusive memristor: properties of basic electrical circuits,'' \emph{European Journal of physics}, vol.~30, no.~4, p. 661, 2009.

\bibitem{xia2019memristive}
Q.~Xia and J.~J. Yang, ``Memristive crossbar arrays for brain-inspired computing,'' \emph{Nature materials}, vol.~18, no.~4, pp. 309--323, 2019.

\bibitem{chua1971memristor}
L.~Chua, ``Memristor-the missing circuit element,'' \emph{IEEE Transactions on circuit theory}, vol.~18, no.~5, pp. 507--519, 1971.

\bibitem{strukov2008missing}
D.~B. Strukov, G.~S. Snider, D.~R. Stewart, and R.~S. Williams, ``The missing memristor found,'' \emph{nature}, vol. 453, no. 7191, pp. 80--83, 2008.

\bibitem{yakopcic2016memristor}
C.~Yakopcic, M.~Z. Alom, and T.~M. Taha, ``Memristor crossbar deep network implementation based on a convolutional neural network,'' in \emph{2016 International joint conference on neural networks (IJCNN)}.\hskip 1em plus 0.5em minus 0.4em\relax IEEE, 2016, pp. 963--970.

\bibitem{yakopcic2017extremely}
------, ``Extremely parallel memristor crossbar architecture for convolutional neural network implementation,'' in \emph{2017 International Joint Conference on Neural Networks (IJCNN)}.\hskip 1em plus 0.5em minus 0.4em\relax IEEE, 2017, pp. 1696--1703.

\bibitem{wen2020ckfo}
S.~Wen, J.~Chen, Y.~Wu, Z.~Yan, Y.~Cao, Y.~Yang, and T.~Huang, ``Ckfo: Convolution kernel first operated algorithm with applications in memristor-based convolutional neural network,'' \emph{IEEE Transactions on Computer-Aided Design of Integrated Circuits and Systems}, vol.~40, no.~8, pp. 1640--1647, 2020.

\bibitem{xia2017mnsim}
L.~Xia, B.~Li, T.~Tang, P.~Gu, P.-Y. Chen, S.~Yu, Y.~Cao, Y.~Wang, Y.~Xie, and H.~Yang, ``Mnsim: Simulation platform for memristor-based neuromorphic computing system,'' \emph{IEEE Transactions on Computer-Aided Design of Integrated Circuits and Systems}, vol.~37, no.~5, pp. 1009--1022, 2017.

\bibitem{wen2019memristor}
S.~Wen, H.~Wei, Z.~Yan, Z.~Guo, Y.~Yang, T.~Huang, and Y.~Chen, ``Memristor-based design of sparse compact convolutional neural network,'' \emph{IEEE Transactions on Network Science and Engineering}, vol.~7, no.~3, pp. 1431--1440, 2019.

\bibitem{ran2020memristor}
H.~Ran, S.~Wen, S.~Wang, Y.~Cao, P.~Zhou, and T.~Huang, ``Memristor-based edge computing of shufflenetv2 for image classification,'' \emph{IEEE Transactions on Computer-Aided Design of Integrated Circuits and Systems}, vol.~40, no.~8, pp. 1701--1710, 2020.

\bibitem{yang2022full}
C.~Yang, X.~Wang, and Z.~Zeng, ``Full-circuit implementation of transformer network based on memristor,'' \emph{IEEE Transactions on Circuits and Systems I: Regular Papers}, vol.~69, no.~4, pp. 1395--1407, 2022.

\bibitem{chen2021automatic}
R.~Chen, W.~Zeng, W.~Fan, F.~Lai, Y.~Chen, X.~Lin, L.~Tang, W.~Ouyang, Z.~Liu, and X.~Luo, ``Automatic recognition of ocular surface diseases on smartphone images using densely connected convolutional networks,'' in \emph{2021 43rd Annual International Conference of the IEEE Engineering in Medicine \& Biology Society (EMBC)}.\hskip 1em plus 0.5em minus 0.4em\relax IEEE, 2021, pp. 2786--2789.

\bibitem{chollet2017xception}
F.~Chollet, ``Xception: Deep learning with depthwise separable convolutions,'' in \emph{Proceedings of the IEEE conference on computer vision and pattern recognition}, 2017, pp. 1251--1258.

\bibitem{hua2018pointwise}
B.-S. Hua, M.-K. Tran, and S.-K. Yeung, ``Pointwise convolutional neural networks,'' in \emph{Proceedings of the IEEE conference on computer vision and pattern recognition}, 2018, pp. 984--993.

\bibitem{li2021native}
B.~Li and G.~Shi, ``A native spice implementation of memristor models for simulation of neuromorphic analog signal processing circuits,'' \emph{ACM Transactions on Design Automation of Electronic Systems (TODAES)}, vol.~27, no.~1, pp. 1--24, 2021.

\bibitem{ioffe2015batch}
S.~Ioffe and C.~Szegedy, ``Batch normalization: Accelerating deep network training by reducing internal covariate shift,'' in \emph{International conference on machine learning}.\hskip 1em plus 0.5em minus 0.4em\relax pmlr, 2015, pp. 448--456.

\bibitem{priyanka2019cmos}
P.~Priyanka, G.~Nisarga, and S.~Raghuram, ``Cmos implementations of rectified linear activation function,'' in \emph{VLSI Design and Test: 22nd International Symposium, VDAT 2018, Madurai, India, June 28-30, 2018, Revised Selected Papers 22}.\hskip 1em plus 0.5em minus 0.4em\relax Springer, 2019, pp. 121--129.

\bibitem{li2024framework}
J.~Li, Y.~Fu, C.-W. Sham, and S.~L. Ma, ``A framework for mapping convolutional neural network onto memristor crossbars,'' in \emph{2024 IEEE 13th Global Conference on Consumer Electronics (GCCE)}.\hskip 1em plus 0.5em minus 0.4em\relax IEEE, 2024, pp. 1390--1393.

\bibitem{sun2018xnor}
X.~Sun, S.~Yin, X.~Peng, R.~Liu, J.-s. Seo, and S.~Yu, ``Xnor-rram: A scalable and parallel resistive synaptic architecture for binary neural networks,'' in \emph{2018 Design, Automation \& Test in Europe Conference \& Exhibition (DATE)}.\hskip 1em plus 0.5em minus 0.4em\relax IEEE, 2018, pp. 1423--1428.

\bibitem{xie202116}
S.~Xie, C.~Ni, A.~Sayal, P.~Jain, F.~Hamzaoglu, and J.~P. Kulkarni, ``16.2 edram-cim: Compute-in-memory design with reconfigurable embedded-dynamic-memory array realizing adaptive data converters and charge-domain computing,'' in \emph{2021 IEEE International Solid-State Circuits Conference (ISSCC)}, vol.~64.\hskip 1em plus 0.5em minus 0.4em\relax IEEE, 2021, pp. 248--250.

\bibitem{li2023adc}
Y.~Li, J.~Chen, L.~Wang, W.~Zhang, Z.~Guo, J.~Wang, Y.~Han, Z.~Li, F.~Wang, C.~Dou \emph{et~al.}, ``An adc-less rram-based computing-in-memory macro with binary cnn for efficient edge ai,'' \emph{IEEE Transactions on Circuits and Systems II: Express Briefs}, 2023.

\bibitem{xiao2023efficient}
H.~Xiao, X.~Hu, T.~Gao, Y.~Zhou, S.~Duan, and Y.~Chen, ``Efficient low-bit neural network with memristor-based reconfigurable circuits,'' \emph{IEEE Transactions on Circuits and Systems II: Express Briefs}, 2023.

\bibitem{hosticka1989very}
B.~Hosticka, R.~Klinke, and H.-J. Pfleiderer, ``A very high slew-rate cmos operational amplifier,'' 1989.

\end{thebibliography}

\vspace{12pt}

\end{document}